\documentclass[acmsmall]{acmart}

\AtBeginDocument{%
  }

\setcopyright{acmcopyright}
\copyrightyear{2025}
\acmYear{2025}
\acmDOI{XXXXXXX.XXXXXXX}

\usepackage{color}
\usepackage{todonotes}
\usepackage{siunitx} % \num
\usepackage{footmisc}
\let\oldtodo\todo
\renewcommand{\todo}[1]{
	\oldtodo[inline]{#1}
}

\usepackage{hyphenat}
\usepackage{tikz}

\newcommand{\code}[1]{\texttt{#1}}

\usepackage{listings}
\acmJournal{JACM}
\acmVolume{37}
\acmNumber{4}
\acmArticle{111}
\acmMonth{8}

\usepackage{amsthm}
\theoremstyle{definition}
\newtheoremstyle{remarkstyle} % https://latex.org/forum/viewtopic.php?t=18631
  {}{}{}{}{\bfseries}{.}{.5em}{{\thmname{#1 }}{\thmnumber{#2}}{\thmnote{ (#3)}}}
\theoremstyle{remarkstyle}

\usepackage{color}
\definecolor{colUniBwOr}{rgb}{0.929,0.431,0.0} % Unibw Orange

\begin{document}

%%
%% The "title" command has an optional parameter,
%% allowing the author to define a "short title" to be used in page headers.
\title{Trilinos: Enabling Scientific Computing across Diverse Hardware Architectures at Scale}

%%
%% The "author" command and its associated commands are used to define
%% the authors and their affiliations.
%% Of note is the shared affiliation of the first two authors, and the
%% "authornote" and "authornotemark" commands
%% used to denote shared contribution to the research.

\author{Matthias Mayr}
\authornote{These authors have contributed equally to this manuscript and, thus, share the first authorship.}
\authornote{Data Science \& Computing Lab, Institute for Mathematics and Computer-Based Simulation, Universit\"at der Bundeswehr M\"unchen, Werner-Heisenberg-Weg 39, 85577 Neubiberg, Germany}
\email{matthias.mayr@unibw.de}
\orcid{0000-0002-2780-1233}

\author{Alexander Heinlein}
\authornotemark[1]
\authornote{Corresponding author}
\authornote{Delft University of Technology, Mekelweg 4, 2628CD Delft, Netherlands}
\email{a.heinlein@tudelft.nl}
\orcid{0000-0003-1578-8104}

\author{Christian Glusa}
\authornotemark[1]
\authornote{Sandia National Laboratories, 1515 Eubank Blvd SE, Albuquerque, NM 87123, United States}
\email{caglusa@sandia.gov}
\orcid{0000-0003-2247-1914}

\author{Sivasankaran Rajamanickam}
\authornotemark[1]
\authornotemark[5]
\email{srajama@sandia.gov}
\orcid{0000-0002-5854-409X}

\author{Maarten Arnst}
\email{maarten.arnst@uliege.be}
\authornote{University of Liege, Quartier Polytech 1, allée de la Découverte 9, 4000 Liege, Belgium}
\author{Roscoe A. Bartlett}\authornotemark[5]
\email{rabartl@sandia.gov}
\orcid{0000-0002-3831-8060}
\author{Luc Berger-Vergiat}\authornotemark[5]
\email{lberge@sandia.gov}
\orcid{0000-0001-5550-3527}
\author{Erik G. Boman}\authornotemark[5]
\email{egboman@sandia.gov}
\author{Karen Devine}\authornotemark[5]\authornote{retired}
\email{devine.hpc@gmail.com}
\author{Graham Harper}\authornotemark[5]
\email{gbharpe@sandia.gov}
\author{Michael Heroux}\authornote{ParaTools, Inc., 18125 Kreigle Lake, Road, Avon, MN 56310, United States}
\email{mheroux@paratools.com}
\orcid{0000-0002-5893-0273}
\author{Mark Hoemmen}\authornote{NVIDIA, 2788 San Tomas Expressway, Santa Clara, CA 95051, United States}
\email{mhoemmen@nvidia.com}
\author{Jonathan Hu}\authornotemark[5]
\email{jhu@sandia.gov}
\author{Brian Kelley}\authornotemark[5]
\email{bmkelle@sandia.gov}
\orcid{0000-0003-3607-360X}
\author{Kyungjoo Kim}\authornotemark[8]
\email{kyungjook@nvidia.com}
\author{Drew P. Kouri}\authornotemark[5]
\email{dpkouri@sandia.gov}
\author{Paul Kuberry}\authornotemark[5]
\email{pakuber@sandia.gov}
\orcid{0000-0002-2426-4591}
\author{Kim Liegeois}
\email{kim.liegeois@amd.com}
\orcid{0000-0002-1182-4078}
\authornote{Advanced Micro Devices, Inc., 2485 Augustine Dr, Santa Clara, CA 95054, United States}
\author{Curtis C. Ober}\authornotemark[5]
\email{ccober@sandia.gov}
\author{Roger Pawlowski}\authornotemark[5]
\email{rppawlo@sandia.gov}
\author{Carl Pearson}\authornotemark[5]
\email{cwpears@sandia.gov}
\orcid{0000-0001-6481-970X}
\author{Mauro Perego}\authornotemark[5]
\email{mperego@sandia.gov}
\author{Eric Phipps}\authornotemark[5]
\email{etphipp@sandia.gov}
\author{Denis Ridzal}\authornotemark[5]
\email{dridzal@sandia.gov}
\author{Nathan V. Roberts}\authornotemark[5]
\email{nvrober@sandia.gov}
\orcid{0000-0003-1536-0749}
\author{Christopher M. Siefert}\authornotemark[5]
\email{csiefer@sandia.gov}
\author{Heidi K. Thornquist}\authornotemark[5]
\email{hkthorn@sandia.gov}
\author{Romin Tomasetti}\authornotemark[6]
\email{romin.tomasetti@uliege.be}
\author{Christian R. Trott}\authornotemark[5]
\email{crtrott@sandia.gov}
\author{Raymond S. Tuminaro}\authornotemark[5]
\email{rstumin@sandia.gov}
\author{James M. Willenbring}\authornotemark[5]
\email{jmwille@sandia.gov}
\orcid{0000-0002-0418-9264}
\author{Michael M. Wolf}\authornotemark[5]
\email{mmwolf@sandia.gov}
\author{Ichitaro Yamazaki}\authornotemark[5]
\email{iyamaza@sandia.gov}

%%
%% By default, the full list of authors will be used in the page
%% headers. Often, this list is too long, and will overlap
%% other information printed in the page headers. This command allows
%% the author to define a more concise list
%% of authors' names for this purpose.
\renewcommand{\shortauthors}{Mayr et al.}

%%
%% The abstract is a short summary of the work to be presented in the
%% article.
\begin{abstract}
  Trilinos is a community-developed, open-source software framework that facilitates building large-scale, complex, multiscale, multiphysics simulation code bases for scientific and engineering problems. Since the Trilinos framework has undergone substantial changes to support new applications and new hardware architectures, this document is an update to ``An Overview of the Trilinos project'' by Heroux et al. (ACM Transactions on Mathematical Software, 31(3):397--423, 2005). It describes the design of Trilinos, introduces its new organization in product areas, and highlights established and new features available in Trilinos. Particular focus is put on the modernized software stack based on the Kokkos ecosystem to deliver performance portability across heterogeneous hardware architectures. This paper also outlines the organization of the Trilinos community and the contribution model to help onboard interested users and contributors.
\end{abstract}

%%
%% The code below is generated by the tool at http://dl.acm.org/ccs.cfm.
%% Please copy and paste the code instead of the example below.
%%
%\begin{CCSXML}
%<ccs2012>
% <concept>
%  <concept_id>10010520.10010553.10010562</concept_id>
%  <concept_desc>Computer systems organization~Embedded systems</concept_desc>
%  <concept_significance>500</concept_significance>
% </concept>
% <concept>
%  <concept_id>10010520.10010575.10010755</concept_id>
%  <concept_desc>Computer systems organization~Redundancy</concept_desc>
%  <concept_significance>300</concept_significance>
% </concept>
% <concept>
%  <concept_id>10010520.10010553.10010554</concept_id>
%  <concept_desc>Computer systems organization~Robotics</concept_desc>
%  <concept_significance>100</concept_significance>
% </concept>
% <concept>
%  <concept_id>10003033.10003083.10003095</concept_id>
%  <concept_desc>Networks~Network reliability</concept_desc>
%  <concept_significance>100</concept_significance>
% </concept>
%</ccs2012>
%\end{CCSXML}

%\ccsdesc[500]{Computer systems organization~Embedded systems}
%\ccsdesc[300]{Computer systems organization~Redundancy}
%\ccsdesc{Computer systems organization~Robotics}
%\ccsdesc[100]{Networks~Network reliability}

%%
%% Keywords. The author(s) should pick words that accurately describe
%% the work being presented. Separate the keywords with commas.

\received{\today}
%\received[revised]{12 March 2009}
%\received[accepted]{5 June 2009}

%%
%% This command processes the author and affiliation and title
%% information and builds the first part of the formatted document.
\maketitle

\section{Introduction}
% !TEX root = main.tex

Trilinos is a community-driven open-source C++ software framework and collection of reusable scientific libraries designed to enable the development of scalable, high-performance algorithms for solving complex, multiscale, and multiphysics engineering and scientific problems on advanced computing architectures.
While Trilinos can run on a variety of hardware platforms ranging from small workstations to large supercomputers, the typical use of Trilinos is on leadership-class systems with new or emerging hardware architectures.

Trilinos was originally conceived as a framework of three packages for distributed memory systems. The original Trilinos publication~\cite{Heroux2005a} described the motivation, philosophy, and capabilities of Trilinos at that time. Seven years later, a two-part collection of 16 papers provided a new overview of Trilinos and many of its packages~\cite{Heroux2012,HerouxIntro2012part1,HerouxIntro2012part2,pawlowski2012automating,Oldfield2012,bochev2012,Boman2012,Baker2012,Kokkos2012,pawlowski2012automatingpart2,Spotz2012,Long2012,Morris2012,Howle2012,Bavier2012a,Gaidamour2012} and described the intervening expansion of Trilinos' capabilities and strategic goals for Trilinos. Trilinos today hews to the overall vision from two decades ago, but has evolved alongside changes in programming models, application needs, hardware architectures, and algorithms. Trilinos has grown from a library of three packages to more than 35 packages with functionality and features supporting a wide range of applications.

For much of its history, packages have relied heavily on the linear algebra classes provided by Epetra.
Epetra did not support larger problems (2B+ unknowns), nor did it allow for accelerator offload.
Since the packages of the old Epetra stack were archived and removed from the main Trilinos repository on GitHub by the end of 2025, this paper describes only the modern Trilinos software stack, which builds upon Tpetra and the Kokkos ecosystem to achieve performance portability across hardware architectures.

This article attempts to capture a snapshot the current state of Trilinos, as an update to the articles published 14 and 21 years ago.
Therefore it focuses on the major developments within Trilinos in the last decade as well as new features and functionality that have been added to advance scientific and engineering applications.
It will give only an overview of the features, and we refer to the extensive list of references for the details of these features.
We are also cognizant of the fact that this article describes software that is being actively developed and constantly evolving.
Hence, we will focus on the high-level features and concepts that we expect to remain stable for several years.
We also briefly touch upon the Trilinos community and software engineering issues with respect to Trilinos.

\subsection{Modern Trilinos:  Performance Portability through Kokkos}

A key goal of modern Trilinos is to offer a performance\hyp{}portable collection of reusable scientific libraries, allowing users to develop applications that achieve high efficiency across all modern high performance computing (HPC) hardware architectures.
This goal emerged when the HPC architecture landscape started diversifying with the
introduction of GPU acceleration for scientific software. Today, nine of the top ten systems in the Top500\footnote{https://top500.org, 66th edition from Nov. 17, 2025} use GPUs, with only a single system using CPUs only. A challenge in this shift is the diversification of vendor-provided programming models: the aforementioned systems have four different preferred programming models: CUDA, OpenACC, HIP and SYCL for the GPU-based systems and OpenMP for the CPU system.

To avoid the necessity of diverging code paths, Trilinos leverages the Kokkos ecosystem\footnote{\url{https://kokkos.org}}~\cite{trott2021kokkos} to write performance portable code. The Kokkos ecosystem originated from the native Trilinos packages Kokkos~\cite{heroux2011toward} and Kokkos Kernels but split off a decade ago into a standalone project hosted and developed independently of Trilinos\footnote{https://github.com/kokkos}.
Now, Trilinos provides snapshots of the two primary Kokkos subprojects (Core and Kernels) that reflect the latest release of Kokkos. Trilinos can also be built against version-compatible external installations of Kokkos --- a capability required for interoperability with other Kokkos-based libraries.

Kokkos enables Trilinos developers to write single source implementations of their packages that perform well on all major HPC hardware architectures. Some of its design principles are also reflected in Trilinos API designs. In particular, the use of \texttt{execution spaces} (where does code run) and \texttt{memory spaces} (where does data reside) as parameters of data structures and algorithms is a common thread throughout Trilinos. This design characteristic enables Trilinos users to manage the heterogeneity of modern architectures and its impact on Trilinos data structures.

\subsection{Trilinos Functionality}

The features and capabilities of Trilinos are divided into several software units called \textit{packages}.
Each Trilinos package has a well-defined set of unique capabilities that is important for scientific or engineering applications. Packages are semi-autonomous, often having their own development team, users, and set of requirements and development principles.  However, packages also follow a general set of Trilinos expectations such as having a designated point-of-contact and following software quality expectations (e.g., sufficient documentation, continuous integration testing, clearly defined dependencies, and using the Trilinos infrastructure for building and installation).

While most packages are developed natively in the Trilinos repository, some packages are developed externally and ``snapshotted'' into Trilinos,
i.e., incorporated as fixed copies of their source code at a specific point in time.
These are (in alphabetical order):
Compadre\footnote{\url{https://github.com/sandialabs/compadre}},
Kokkos and Kokkos Kernels,
Krino,
ROL\footnote{\url{https://github.com/sandialabs/rol}},
SEACAS\footnote{\url{https://github.com/sandialabs/seacas}},
Sierra ToolKit (STK),
and TriBITS\footnote{\url{https://github.com/TriBITSPub/TriBITS}}.
Some reasons for snapshotting external packages into Trilinos include:
external packages and repositories facilitate easy and independent usage by other software projects without having to pull in Trilinos;
snapshotting provides an integration process for external packages into Trilinos, ensuring continual compatibility between all packages;
and snapshotting allows easy integration into Trilinos-based application codes, alleviating users from managing these dependencies themselves.
We refer to the Trilinos website (\url{https://trilinos.github.io}) for a brief description of these packages.

In the following sections of this paper, we group the Trilinos packages that share common objectives (e.g., solving linear systems) together into Trilinos' five \textit{product areas}:  Core, Linear Solvers and Preconditioners, Nonlinear Solvers and Analysis Tools, Discretization Tools, and Framework.
Trilinos' packages and their assignment to a product area are illustrated in Figure~\ref{fig:GraphicalOverview}.
\begin{figure}
\centering
\input{graphical_overview.tex}
\caption{Organization of the Trilinos library into five product areas and their packages: Product areas are labeled in bold font with their respective packages.
Snapshotted packages described in this paper are marked by dashed boxes.}
\label{fig:GraphicalOverview}
\end{figure}
We briefly describe these product areas below.

\paragraph{Core} Core packages cover all aspects of creating, distributing, and mapping data to processing elements (cores, threads, nodes), load balancing, and redistributing data. This includes Trilinos' abstractions for linear algebra data structures and algorithms as well as concrete implementations such as Tpetra linear algebra data structures. On a modern accelerator-based compute node, the abstractions provided by the Kokkos library become critical for Tpetra. These capabilities are described in detail in Section~\ref{sec:data_services}.

\paragraph{Linear Solvers and Preconditioners} The wide variety of applications that use Trilinos need a diverse set of linear solvers. Trilinos has support for both iterative and direct linear solvers, including interfaces to external solver packages. There are a number of preconditioner options from multithreaded or performance portable node-level preconditioners to scalable multilevel domain decomposition or multigrid preconditioners. The preconditioners and solvers use the data abstractions from the core packages. Section \ref{sec:lin_solve} provides a detailed description of these features.

\paragraph{Nonlinear Solvers and Analysis Tools} These packages provide algorithms for computational simulations and design. Capabilities include solvers for nonlinear equations, parameter continuation, bifurcation tracking, optimization, and uncertainty quantification. Trilinos also provides utility packages to evaluate quantities of interest required by the analysis algorithms. Additional capabilities include automatic differentiation technology to evaluate derivatives and embedded ensemble propagation for uncertainty quantification. These packages and their capabilities will be discussed further in Section~\ref{sec:nonlin_solve}.

\paragraph{Discretization Tools} This collection of packages provides functionality for the discretization of differential equations. In particular, it supports mesh-free and mesh-based spatial discretizations, with a focus on high-order finite elements, and time integration. Discretization tools also include support for algorithmic differentiation and for managing directed acyclic graphs of evaluation kernels. These capabilities are described further in Section~\ref{sec:discretization} in detail.

\paragraph{Framework} The Framework product is different than the other Trilinos products in that most of the resources and services are not associated with Trilinos packages. The Framework product area is primarily focused on activities such as developing and maintaining infrastructure for automated testing and documentation, as well as associated workflows. A small number of infrastructure and cross-cutting packages are also associated with the Framework, for example the Python wrapper package PyTrilinos2. Section~\ref{sec:framework} provides further details.

\subsection{Outline of this manuscript}

Sections~\ref{sec:data_services} -- \ref{sec:framework} provide details on Trilinos' product areas.
Section~\ref{sec:community} briefly touches upon the Trilinos community, software engineering aspects as well as the integration of Trilinos into the landscape of scientific software frameworks.
Finally, Section~\ref{sec:conclusion} concludes the manuscript.

In comparison to the first Trilinos publication from 2005~\cite{Heroux2005a}
and its updates from 2012~\cite{Heroux2012,HerouxIntro2012part1,HerouxIntro2012part2,pawlowski2012automating,Oldfield2012,bochev2012,Boman2012,Baker2012,Kokkos2012,pawlowski2012automatingpart2,Spotz2012,Long2012,Morris2012,Howle2012,Bavier2012a,Gaidamour2012},
the present manuscript not only provides updates on newly added capabilities,
but more importantly reflects the overhaul of Trilinos towards performance portability through the full integration of the Kokkos ecosystem~\cite{edwards2014,trott2021kokkos,trott2022kokkos}.
Many packages described in the original Trilinos publication~\cite{Heroux2005a} have been replaced by Kokkos-capable successor implementations,
among them (in alphabetical order)
Amesos2 replacing Amesos,
Belos replacing AztecOO,
Ifpack2 replacing Ifpack,
MueLu replacing ML,
and Tpetra replacing Epetra.
The present manuscript introduces these new packages and their capabilities and outlines their interplay with Kokkos.

\section{Trilinos Core}
\label{sec:data_services}
% !TEX root = main.tex

The Core product area provides essential tools for managing and distributing data across processing elements in parallel computing environments, addressing challenges such as load balancing and data redistribution. Key components discussed here include the Kokkos performance portability framework, Teuchos tools, Tpetra distributed-memory linear algebra, Thyra abstractions for high-level formulations of numerical algorithms, and Zoltan2 for combinatorial algorithms in parallel computing.

\subsection{Performance Portability: Kokkos Core}\label{subsec:kokkos}
Kokkos Core\footnote{For historical reasons the package name is ``Kokkos'' since it precedes the creation of other Kokkos subprojects such as Kokkos Kernels.} is a programming model designed for performance portability across hardware architectures such as CPUs and GPUs.
While the development of Kokkos Core has originated as a package of Trilinos, Kokkos has become an independent project in 2015
and is nowadays snapshotted into Trilinos for the pure convenience of dependency management.
Two primary insights into achieving performance portability shaped Kokkos' development:
i) abstractions for both parallel execution and data structures are required, and
ii) the fundamental programming model must be descriptive rather than prescriptive.

The Kokkos programming model includes \texttt{execution spaces} to specify where an algorithm should run, \texttt{execution policies} to indicate how the algorithm should be run in terms of hardware resources, and \texttt{execution patterns} to designate whether to execute a \mbox{\texttt{parallel\_for}}, \mbox{\texttt{parallel\_reduce}}, or \mbox{\texttt{parallel\_scan}}. The data structure abstractions are built around the concepts of \texttt{memory spaces} to indicate where memory is allocated, \texttt{memory layouts} to specify how data is laid out in memory, and \texttt{memory traits} that enable customizations of how data is accessed. The primary data structure incorporating
these concepts is \texttt{Kokkos::View}, a multidimensional array abstraction that inspired
the ISO C++23 \texttt{mdspan} class\footnote{\url{https://www.open-std.org/jtc1/sc22/wg21/docs/papers/2022/p0009r18.html}}. Detailed descriptions of Kokkos Core can be found
in~\cite{edwards2014} and \cite{trott2022kokkos}.
Development of Kokkos Core itself is carried out in its own GitHub repository\footnote{https://github.com/kokkos/kokkos}.

\subsection{Performance Portable Kernels: Kokkos Kernels}\label{subsec:kk}
Kokkos Kernels~\cite{rajamanickam2021kokkoskernels} is part of the Kokkos ecosystem~\cite{trott2021kokkos} and provides node-local implementations of mathematical kernels widely used across packages in Trilinos. As a member of the Kokkos ecosystem, Kokkos Kernels tightly integrates Kokkos features and aims at delivering performance\hyp{}portable algorithms across major CPU- and GPU-based HPC systems. Due to its node-local nature, Kokkos Kernels does not rely on MPI or other communication libraries, unlike numerous other packages in Trilinos.

The implementations of Kokkos Kernels algorithms leverage the hierarchical parallelism exposed by the Kokkos library~\cite{kim2017designing}. To ensure flexibility for the distributed libraries that might call its algorithms, Kokkos Kernels provides thread-safe implementations for most of its kernels, with increasing coverage for asynchronous implementations that allow an execution space instance to be passed when calling a kernel. Kokkos Kernels also serves as a major point of integration for vendor optimized libraries such as cuBLAS, cuSPARSE, rocBLAS, rocSPARSE, MKL, ARMpl and others.

The capabilities provided by Kokkos Kernels can be divided into four major categories:
1. BLAS algorithms, 2. sparse linear algebra and preconditioners, 3. graph algorithms, and
4. batched dense and sparse linear algebra~\cite{liegeois2023performance,10.1145/3431921}. The main
points of integration of Kokkos Kernels in Trilinos are Tpetra for the dense and sparse
linear algebra capabilities, Ifpack2 for the preconditioners and batched algorithms,
and the multigrid package MueLu which relies directly and indirectly on BLAS and graph algorithms and specialized implementations of graph coloring/coarsening~\cite{kelley2022parallel} and fused Jacobi-SpGEMM (generalized sparse matrix-matrix multiply) kernels.
Additionally, batched dense factorization algorithms are used in the construction of grid transfers.

Similar to the Kokkos library, Kokkos Kernels is developed in its own GitHub
repository\footnote{https://github.com/kokkos/kokkos-kernels} outside of the Trilinos
GitHub repository. Every version of the library is integrated and tested in Trilinos
as part of the Kokkos ecosystem release process. Additional information on Kokkos
Kernels' capabilities can be found in~\cite{deveci2018multithreaded,wolf2017fast}.

\subsection{Tools: Teuchos}

Teuchos provides a collection of helper and utility components designed to simplify common programming tasks and to ensure a consistent coding style across Trilinos packages.
These tools include memory management classes~\cite{bartlett2010} such as smart pointers and arrays,
parameter lists for communicating hierarchical lists of parameters between library or application layers,
templated wrappers for BLAS and LAPACK, XML parsers, MPI, and other supporting functionality.
Teuchos thereby provides a unified ``look and feel'' across Trilinos packages and helps to avoid common programming mistakes.
It offers reusable building blocks for such recurring tasks implemented in a consistent and coherent manner to be used throughout the entire code base.

\subsection{Distributed-Memory Linear Algebra: Tpetra}\label{subsec:tpetra}
Tpetra~\cite{Baker2012,hoemmen2015tpetra} provides the distributed-memory
infrastructure for linear algebra objects such as sparse graphs,
sparse matrices, and dense vectors. These linear algebra objects are
implemented using MPI-based abstractions for the inter-process communication,
and Kokkos-based data structures for the portions of the object
owned by each MPI process. Kokkos Kernels is used for
performing local, parallel computations within each MPI process.
Due to the fundamental nature of MPI distributed data structures, the majority of Trilinos packages depend directly or indirectly on Tpetra.

The objects provided are templated on scalar type (e.g., \code{double}, \code{float}, or a Sacado or Stokhos type), local index type, global index type, and Kokkos device (pair of execution and memory spaces).
The distribution of the linear algebra objects is described by a \code{Map} data structure,
which defines which elements are assigned to which MPI ranks.
Here ``element'' could be a row or column of a sparse matrix or an entry of a vector.
Global indices number elements across an entire distributed object, while local indices number elements only within a rank.
The \code{Map} encodes the relation between local indices and global indices.
An illustration for a vector, that is distributed to two MPI processes, is given in Figure~\ref{fig:Map}.
Since memory capacity limits the sizes of local objects, local indices can be safely represented with a more compact type than global indices (e.g., 32-bit \code{int} instead of 64-bit \code{long}).

\begin{figure}
% !TEX root = main.tex

\begin{tikzpicture}

\definecolor{colUniBwOr}{rgb}{0.929,0.431,0.0}
\definecolor{colUniBwGr}{RGB}{113,112,114}
\definecolor{procZero}{named}{colUniBwOr}
\definecolor{procOne}{named}{colUniBwGr}

\providecommand{\mapentry}[2]{\colorbox{#2!30}{\makebox[1em][c]#1}}

\def\textPos{1.2}
\def\mapPos{2.4}
\def\procPos{4}

\node at (0,0) {
$
\mathbf{v} =
\begin{bmatrix}
\mapentry{v}{procZero}\\
\mapentry{w}{procOne}\\
\mapentry{x}{procZero}\\
\mapentry{y}{procOne}\\
\mapentry{z}{procZero}
\end{bmatrix}
$
};

\draw [very thick, -{stealth}] (0.8,0.2) -- (2.1,1);
\draw [very thick, -{stealth}] (0.8,-0.2) -- (2.1,-1);
\draw [thick, dashed] (1.5,0) -- (8,0);

\begin{scope}[shift={(3,1)}]
\node at (0,0) {
$
\mathbf{v}_0 =
\begin{bmatrix}
\mapentry{v}{procZero}\\
\mapentry{x}{procZero}\\
\mapentry{z}{procZero}
\end{bmatrix}
$
};

\node at (\textPos,0) {with};
\node at (\mapPos,0) {
$
\mathbf{M}_0 =
\begin{bmatrix}
0\\
2\\
4
\end{bmatrix}
$
};
\node at (\procPos,0) {on process~$0$};

\end{scope}

\begin{scope}[shift={(3,-1)}]
\node at (0,0) {
$
\mathbf{v}_1 =
\begin{bmatrix}
\mapentry{w}{procOne}\\
\mapentry{y}{procOne}
\end{bmatrix}
$
};

\node at (\textPos,0) {with};
\node at (\mapPos,0) {
$
\mathbf{M}_1 =
\begin{bmatrix}
1\\
2
\end{bmatrix}
$
};
\node at (\procPos,0) {on process~$1$};

\end{scope}

\end{tikzpicture}
\caption{Illustration of a \code{Tpetra::Map} for an exemplary vector~$\mathbf{v}\in\mathbb{R}^5$ distributed to two processes~$p\in\{0,1\}$: Entries~$v$, $x$ and~$z$ are stored on process~$0$, entries~$w$ and~$y$ are stored on process~$1$. Local-to-global mapping of indices is given by~$\mathbf{M}_p$, where the $i$th element of~$\mathbf{M}_p$ denotes the unique global index corresponding to the local index~$i$ on rank~$p$. In more complicated scenarios, entries can also be shared between processes.}
\label{fig:Map}
\end{figure}

\code{DistObject} is Tpetra's abstract base class for general distributed objects.
Subclasses define how to pack elements into a buffer for sending with MPI, and how to unpack received elements.
The \code{Import} and \code{Export} classes describe how one \code{DistObject} communicates data to another.
They are both created from a pair of \code{Map}s, a source and a target. For example, halo exchange on a vector is an \code{Import} where
the source is the vector's original distribution, and the target is the same except that each rank
also ``owns'' the degrees-of-freedom in the halo around its subdomain.

The following are the primary linear algebra objects in Tpetra (all are subclasses of \code{DistObject}):
\begin{itemize}
\item \emph{(Multi)Vectors} enable to storage of dense vectors or collections of
vectors (multivectors) and associated BLAS-1 like kernels (e.g., dot
products, norms, scaling, vector addition, pointwise vector
multiplication) as well as tall skinny QR (TSQR) factorization for multivectors.
\item \emph{CrsGraph} implements a sparse graph in compressed row storage (CRS)
format. Graphs also include import/export objects for use in
halo/boundary exchanges associated with the graph.
\item \emph{CrsMatrix} and \emph{BlockCrsMatrix} are sparse matrices in CRS and
block compressed sparse row (BSR) formats. Associated format-specific computational kernels include
sparse matrix-vector product (SPMV), sparse matrix-matrix
multiplication (SPGEMM) and triple-product, sparse matrix-matrix addition, sparse matrix transpose, diagonal extraction,
Frobenius norm calculation, and row/column scaling.
\end{itemize}

\subsection{Abstract Numerical Algorithms: Thyra}
The Thyra package provides abstractions for implementing high-level abstract numerical algorithms (ANAs). Thyra is used by other Trilinos packages for implementing ANAs such as solvers for linear and nonlinear equations, bifurcation and stability analysis, optimization, and uncertainty quantification. The abstractions hide concrete linear algebra and solver implementations so that algorithm development is separated from the implementation of the underlying kernels, allowing the same code to be written once and reused across many application codes. For example, one could write a Krylov solver and, without modifications to computational kernels that are hidden behind the abstraction layer, use it with multiple concrete linear algebra implementations such as Tpetra in Trilinos or the vector/matrix objects in PETSc~\cite{petsc-web-page}.

Thyra is organized into three sets of abstractions depending on the type of operation required. All are templated on the scalar type. These sets are:
\begin{itemize}
\item \emph{Vector/Operator Interfaces:} The first level is an abstraction for vectors and operators \cite{Bartlett2007}. This set includes abstractions for vector spaces, vectors, multivectors, and linear operators. Thyra provides a default set of vector-vector and vector-operator operations for standard linear algebra tasks such as AXPY from BLAS. Defining an extensible interface such that users can add arbitrary vector/matrix operations and maintain performance can be difficult. A general tool to achieve this is the Trilinos utility package RTOp \cite{rtop}, which provides an interface for users to add new Thyra operations. RTOp accepts a vector of Thyra input objects and a vector of Thyra output objects along with a functor. RTOp will then apply the functor to each element of the data.

\item \emph{Operator Solve Interfaces:} The second set of abstractions supplies abstract linear solvers that can be attached to operators. The interfaces include base classes for preconditioners, preconditioner factories, linear operators with a solver, and factories for linear operators with a solver. ``Factories'' build and update the preconditioner and linear solver objects as needed. The solver interfaces allow one to wrap any concrete linear solver or preconditioner for use with the ANAs. As a concrete implementation of this concept, the Stratimikos package wraps all of the Trilinos preconditioners and linear solvers in Thyra abstractions. Run-time configuration of Stratimikos is driven by \code{Teuchos::ParameterList} options.
Details on Stratimikos are given in Section~\ref{sec:Stratimikos}.

\item \emph{Nonlinear Interfaces:} The final set of abstractions supports nonlinear operations. Thyra provides an interface that wraps nonlinear solvers. Implicit time integration and optimization solvers typically require nonlinear solves and can abstract away the concrete implementation via Thyra. The NOX package in Trilinos (described in Section~\ref{sec:nox}) provides a concrete implementation of this interface. The final interface in Thyra is for querying a physics application for common objects required by Newton-based solvers. The \code{Thyra::ModelEvaluator} interface allows codes to ask the application to evaluate residuals, Jacobians, parametric sensitivities, Hessians, and post-processed quantities/derivatives of interest. The interface is designed to be stateless (i.e., consecutive calls with the same inputs should produce the same outputs). This design was chosen to avoid pitfalls in complex application codes interacting with general nonlinear algorithms. The design does not preclude its use in stateful applications that have a ``memory'' between timesteps, but rather forces the explicit use of the stateful information in the interface.
\end{itemize}

\subsection{Combinatorial algorithms with hybrid parallelism: Zoltan2}
Zoltan2 is a package for partitioning, load balancing, and other combinatorial scientific computing algorithms. It can be viewed as a successor to Zoltan~\cite{Boman2012} that provides both tighter integration with other Trilinos packages and support for hybrid parallelism. It is used primarily to partition and distribute data objects (such as sparse matrices) across processors.

Zoltan2 has a highly modular design. The user provides input via input adapters. Adapters for sparse graphs and matrices are provided for Xpetra, which can also be used to interface Tpetra data structures. The models create a graph, hypergraph, or coordinate model from the user input, which is typically a sparse matrix, mesh, or vector. The algorithms then use this model to solve the particular problem (e.g., partitioning or coloring) requested by the user.

Zoltan2 supports the use of both distributed-memory parallelism and on-node parallelism on CPUs and GPUs by building on Kokkos.
Algorithms that support such hybrid parallelism include Multi-jagged~\cite{Z1} (geometric partitioning), Sphynx~\cite{Z2} (spectral graph partitioning), graph coloring~\cite{Z5}, and MPI task placement~\cite{Z3}.

\section{Linear Solvers and Preconditioners}
\label{sec:lin_solve}
% !TEX root = main.tex

Trilinos offers many linear solver capabilities: dense and sparse direct solvers, iterative solvers, shared-memory preconditioners local to a compute node, and scalable distributed memory domain decomposition and multigrid methods. Furthermore, interfaces to several third-party direct solvers are provided. All native solver capabilities are built on top of Kokkos and are GPU capable to varying degrees; any exceptions are noted in the detailed descriptions below.

\subsection{Iterative Linear Solvers and Krylov Methods: Belos}

Belos~\cite{Bavier2012a} provides next-generation iterative linear solvers and a powerful
framework to develop solvers for linear systems and least-squares problems. This
framework is powerful because it provides several algorithmic abstractions
that facilitate code reuse and extensibility. These abstractions are at every level
of the iterative solver hierarchy, from the foundational linear algebra to the solver managers
that embody the overall solution strategy, facilitating flexible solver development.

Belos fundamentally decouples the algorithms from the implementation of the underlying linear algebra object using abstract traits classes that prescribe the required matrix and vector operations. Adapters are necessary to concrete linear algebra implementations, which are provided for Tpetra and Thyra. Users can also implement their own interfaces to leverage existing matrix and vector representations. Next, there are abstractions to orthogonalization to ease the integration of application or architecture-specific orthogonalization methods. Implementations of iterated classical Gram-Schmidt, DGKS-corrected (Daniel, Gragg, Kaufman, and Stewart-corrected) classical Gram-Schmidt, and iterated modified Gram-Schmidt algorithms are provided.  Finally, solver managers encapsulate the strategy for solving a linear system or least-squares problem using abstract interfaces to iteration kernels.  As a result, the algorithms developed using Belos abstractions can be relatively agnostic to data layout in memory or distributed over processors and parallel matrix/vector operations.

Belos supersedes the AztecOO package~\cite{Heroux2004a} by providing solver managers for individual iterative methods. A solver manager wraps a specific iteration and provides the high-level control logic and supporting functionality for one method, such as CG or GMRES, rather than for a collection of methods. Belos includes solver managers for single-vector iterative solvers, as well as extensions of several of these methods to block iterative methods for solving linear systems with multiple right-hand sides.  Single-vector solver managers include conjugate gradient (CG), minimum residual (MINRES), generalized minimal residual (GMRES), stabilized biconjugate gradient (BiCGStab), and transpose-free QMR (TFQMR) as well as ``seed'' solvers (hybrid GMRES, PCPG), subspace recycling solvers (GCRO-DR and RCG), and least-squares solvers (LSQR).  There are versions of CG, GMRES, and GCRO-DR that construct a block Krylov subspace to solve for one or more right-hand sides.  There are also ``pseudo-block'' versions of CG, GMRES, and TFQMR that apply the single-vector iteration simultaneously on multiple right-hand sides, where matrix and preconditioner applications are aggregated to achieve better performance. Any single-vector iteration can be used to solve multiple right-hand sides as well but, if a block or pseudo-block version is not used, the solver will sequentially solve the right-hand sides.

In recent years, application or architecture-specific variants of the classic iterative methods have been integrated into Belos. Some of these variations are so minor that they are enabled through input parameters on the classic method. For instance, flexible GMRES~\cite{Saad1993a} is an option for the block GMRES solver, and pipelined CG~\cite{GHYSELS2014224} is an option for both the block CG and pseudo-block CG solver.  Additionally, there are stand-alone solvers for a pseudo-block stochastic CG method~\cite{Parker2012SamplingGD} and a fixed-point iteration method.

\subsection{One-Level Domain Decomposition and Basic Iterative Methods: Ifpack2}

Trilinos provides domain decomposition approaches in two different
packages: Ifpack2 and ShyLU (specifically the ShyLU\_DD and FROSch
subpackages). Ifpack2 implements overlapping additive Schwarz
approaches with several options for the local subdomain solves. The
local subdomain solvers may be either CPU-only versions of incomplete
factorization preconditioners implemented in Ifpack2 itself, such as
ILU(k) and ILUt (thresholded ILU), or architecture-portable algorithms
for incomplete factorizations and triangular solvers implemented in
Kokkos Kernels. It is also possible to use direct solvers as subdomain
solvers, and ShyLU also supports shared-memory inexact
incomplete factorization preconditioners.
Such one-level preconditioners can be used as smoothers within multigrid
methods, for solving ``simpler'' problems where the setup cost of more robust and scalable multilevel methods is prohibitive in comparison to the reduction in the number of iterations, or when the underlying problem is simply not amenable to multilevel methods.

Ifpack2 also supplies classic iterative methods based on matrix-splitting techniques, such as Jacobi iteration, Gauss-Seidel, and an MPI-oriented hybrid of Jacobi and Gauss-Seidel (e.g., Jacobi between ranks and Gauss-Seidel on them). Ifpack2 also provides preconditioners
based on Chebyshev iterations. The aforementioned preconditioners are available both in point and block
forms and can operate on CRS and BSR matrices. In the block case,
line relaxation is also supported, while in the point case, techniques
like Vanka relaxation \cite{Vanka1986} are possible.  Auxiliary-space
smoothing for Maxwell and Darcy type problems in the style of
Hiptmair \cite{Hiptmair1997} are also supported.
Local kernels are either implemented in Ifpack2 itself
but can also be called from Kokkos Kernels for performance\hyp{}portable shared-memory algorithmic variants of Gauss-Seidel and Jacobi iterations.

\subsection{Multilevel Domain Decomposition Methods: FROSch}
\label{ssec:frosch}

FROSch (Fast and Robust Overlapping Schwarz) is a framework for the construction of multilevel Schwarz domain decomposition preconditioners. Besides \emph{parallel scalability}, FROSch emphasizes \emph{applicability and robustness across a wide range of challenging problems} while supporting \emph{algebraic construction} of the preconditioners, that is, solely from the fully assembled system matrix.

Most FROSch preconditioners can be constructed algebraically, though some variants can take advantage of additional geometric input. The algebraic construction relies on two key algorithmic components: first, the creation of an overlapping domain decomposition at the initial level based on the sparsity pattern of the system matrix, similar to Ifpack2; and second, the integration of extension-based coarse spaces, as in the classical two-level Generalized Dryja--Smith--Widlund (GDSW) preconditioner~\cite{dohrmann_domain_2008} and its variants. While the initial version of FROSch~\cite{heinlein_parallel_2016} was based on the outdated Epetra linear algebra framework, the current implementation~\cite{heinlein_frosch_2020} leverages Xpetra. Originally designed as a lightweight wrapper around Epetra and Tpetra, the Xpetra package facilitated over the years compatibility with both the Epetra and Tpetra stacks. With the deprecation of Epetra, it now exclusively supports the Tpetra stack. Algorithmic variants of Schwarz methods implemented in FROSch include:
\begin{itemize}
	\item \emph{Extension-based coarse spaces based on a partition of unity on the interface}, such as classical GDSW, reduced dimension GDSW (RGDSW) coarse spaces, and multiscale finite element method (MsFEM) coarse spaces; cf.~\cite{heinlein_parallel_2016,heinlein_improving_2018};
	\item \emph{Monolithic Schwarz preconditioners} for block systems; cf.~\cite{heinlein_monolithic_2019}.
	\item \emph{Multilevel Schwarz preconditioners}, which are obtained from two-level Schwarz preconditioners by recursively applying Schwarz preconditioners as an inexact solver for the coarse problems; cf.~\cite{heinlein_parallel_2022}.
\end{itemize}

The wide applicability of FROSch has been demonstrated for various challenging application problems, including: scalar elliptic and elasticity problems~\cite{heinlein_parallel_2016}, possibly with heterogeneities~\cite{alves2024computationalstudyalgebraiccoarse}; computational fluid dynamics problems~\cite{heinlein_monolithic_2019,heinlein_comparison_2025}; time-harmonic Maxwell's and fluid-structure interaction problems~\cite{heinlein2024couplingdealiifroschsustainable}; pharmaco-mechanical interactions in arterial walls~\cite{balzani_computational_nodate}; and coupled multiphysics problems for land ice simulations~\cite{heinlein_frosch_2022}. The latter three have been solved using monolithic preconditioning techniques. To extend robustness for heterogeneous model problems, an implementation of spectral coarse spaces~\cite{heinlein_adaptive_2019} is currently under development. FROSch preconditioners have scaled to more than \num{200000} cores on the Theta Cray XC40 supercomputer at the Argonne Leadership Computing Facility (ALCF); cf.~\cite{heinlein_parallel_2022}.

In its current implementation, FROSch assumes a one-to-one correspondence of subdomains and MPI ranks. Using an interface to the other solver packages in Trilinos, a variety of inexact solvers can be employed for the solution of the local subdomain problems. An extension to multiple subdomains per MPI rank is currently being implemented. Using Kokkos and Kokkos Kernels, FROSch has recently also been ported to GPUs~\cite{yamazaki_experimental_2023} with
performance gains for the triangular solve or inexact solves with ILU on GPUs; speedups of up to $4.8\times$ are observed when comparing ILU solve times on CPU and GPU.

A demo/tutorial for FROSch can be found at the GitHub repository~\cite{frosch_demo}.

\subsection{Multigrid Methods: MueLu}

MueLu is a flexible and scalable high-performance multigrid solver library.
It provides a variety of multigrid algorithms for problems ranging from Poisson-like operators, over elasticity, convection-diffusion, Navier-Stokes, and Maxwell's equations,
all the way to multigrid methods for coupled multiphysics systems.
Besides its strong focus on aggregation-based algebraic multigrid (AMG) methods,
MueLu comes with specialized capabilities for (semi-)structured grids to perform semi-coarsening along grid lines
while forming the coarse operator via a Galerkin product (in contrast to classical geometric multigrid methods).
MueLu is easily extensible which makes it a good framework for the research and development of new multigrid preconditioning methods, e.g. different types of strength-of-connection or drop schemes.
Its weak and strong scalability up to \num{131000} cores of a Cray XC40 and one million cores of a Blue Gene/Q system, even for vector-valued partial differential equations (PDEs) on unstructured meshes, have been shown in~\cite{Lin2017a,Thomas2019a}.

MueLu provides several approaches to constructing and solving the multilevel problem:

\begin{itemize}
	\item \emph{Algebraic smoothed aggregation approach}~\cite{Vanek1996a}:
	      The matrix graph is colored to create aggregates (groups) of nodes.
	      These aggregates define a tentative projection operator.
	      A final projection operator is created by applying a smoother to the tentative operator.
	      There are a variety of deterministic and non-deterministic coloring algorithms implemented directly
	      in MueLu and in Kokkos Kernels.  For a full description, see~\cite{BergerVergiat2023a}.

	\item \emph{Algebraic multigrid for Maxwell's equations}:
	      MueLu implements specialized solvers for the solution of curl-curl problems~\cite{BochevHuEtAl2008_AlgebraicMultigridApproachBased}.
	      Scaling results on Haswell, KNL, ARM and NVIDIA V100 GPUs at full machine scale can be found in~\cite{BettencourtBrownEtAl2021_EmpirePic}.

	\item \emph{Multigrid for multiphysics problems}:
	      MueLu implements a tool box to compile multi-level block preconditioners for block matrices arising from coupled multiphysics problems.
	      Applications range from Navier-Stokes equations
	      over surface-coupling (as in fluid\hyp{}structure interaction or contact mechanics~\cite{Wiesner2021a})
	      to volume-coupled problems (e.g., in magneto-hydro dynamics~\cite{Ohm2022a}).

	\item \emph{Semi-coarsening algebraic multigrid approach}~\cite{Tuminaro2016a}:
	      Specialized aggregation procedures for three-dimensional meshes generated by extrusion of a two-dimensional unstructured grid
	      allow to first coarsen in the direction of extrusion to reduce the system to a two-dimensional representation and then perform classical aggregation-based AMG
	      for the remaining coarsenings.

	\item \emph{AMG for (semi-)structured grids}:
	      MueLu has a structured aggregation capability in which the user may specify the coarsening rate used to compute interpolation operators.
	      The coarse operators are then formed via a Galerkin product to avoid remeshing on the coarse levels.
	      This work has been extended to semi-structured grids to leverage structured-grid computational performance also for globally unstructured grids~\cite{Mayr2022a}.

	\item \emph{Geometric multigrid / matrix-free capabilities}:
	      MueLu treats the objects in its hierarchy as operators instead of matrices whenever possible.
	      This enables one to use MueLu in a matrix-free fashion provided the user can supply their own grid transfers and coarse operators.
\end{itemize}

MueLu has a required dependency on Kokkos,
and the main code paths in MueLu (e.g., smoothed aggregation, semi-coarsening, $p$-coarsening, structured aggregation)
utilize Kokkos and have been shown to scale on device \cite{BettencourtBrownEtAl2021_EmpirePic}.
Some of the earlier algorithms such as energy minimization~\cite{Sala2008a} and the variable degree-of-freedom Laplacian have not been adapted to fully utilize Kokkos yet,
however, they are part of ongoing efforts to use Kokkos throughout MueLu completely and to merge serial and device-capable algorithms.

There are a variety of resources to learn more about MueLu.
An overview is given on the MueLu website\footnote{\url{https://trilinos.github.io/muelu.html}}.
The MueLu User's Guide~\cite{BergerVergiat2023a} summarizes installation instructions and a reference to most of MueLu's configuration parameters.
The MueLu Tutorial~\cite{Mayr2023b} introduces beginners and experts to various topics in MueLu and shows how to solve or precondition different linear systems using MueLu.
Details on the compatibility of MueLu and its predecessor ML~\cite{Heroux2005a,Gee2006a} can be found in the MueLu User's Guide~\cite{BergerVergiat2023a}.

Besides its C++ API, MueLu offers a MATLAB interface, MueMex, to provide access to MueLu's aggregation and solver routines from MATLAB.
MueMex allows users to setup and solve arbitrarily many problems with either MueLu as a preconditioner, Belos as a solver and Tpetra for data structures.

MueLu requires the packages Teuchos, Tpetra, Xpetra, Kokkos, Kokkos Kernels, Amesos2, Ifpack2, and Zoltan2. Additionally, it supports interfaces to abstraction layer packages such as Stratimikos and Thyra through the MueLu adapters library. These interfaces are required to use MueLu with the Teko block preconditioning package. For more details on using MueLu with Teko, see the MueLu examples directory in Trilinos.

\subsection{Direct Linear Solvers: Amesos2 and ShyLU}

Amesos2~\cite{Bavier2012a} serves as the central collection of direct solvers for any type of linear systems:
it natively implements the direct solver KLU2 \cite{Davis2010a},
it offers a uniform interface to third-party direct linear solvers such as CHOLMOD, MUMPS, Pardiso\_MKL, SuperLU, SuperLU\_MT, SuperLU\_Dist, and STRUMPACK,
and it provides an interface to the on-node sparse direct solvers implemented in ShyLU.

ShyLU~\cite{ShyLUCore2014} provides an implementation of three solvers: Basker, Tacho, and a distributed-memory solver based on the Schur complement method, which all have been developed for specific target applications.

Basker~\cite{Basker2017} is a shared-memory sparse direct solver based on LU factorization for the problems that have the block triangular form (BTF) typically seen in circuit simulation applications. Basker uses this structure to factor and solve the diagonal blocks in parallel. The larger diagonal blocks can themselves be factored in parallel by discovering the parallelism available using a nested-dissection reordering. Basker focuses on exploiting thread-parallelism on the multi-core CPU architectures.
Amesos2 also has a templated implementation of a sequential KLU solver called KLU2, which also exploits the BTF structure.

Tacho is a shared-memory sparse direct solver that exploits supernodal block structure commonly found in sparse direct factorizations of matrices from mechanics applications. Tacho exploits this supernodal structure for both factorization and triangular-solve phases. It is based on Kokkos. Originally, Tacho implemented a task-parallel Cholesky factorization of sparse symmetric positive definite (SPD) matrices~\cite{Tacho2018}. However, to improve its portability, it has been extended to compute the sparse factorization based on level-set scheduling. Moreover, its functionality has been extended to the computation of an LDLT factorization of symmetric indefinite matrices, as well as the computation of an LU factorization of general matrices with a symmetric sparsity pattern.

In addition to their stand-alone use, the aforementioned node-level solvers may be used as the local solvers for domain decomposition preconditioners (Ifpack2 or FROSch) or as the coarse solvers for multilevel preconditioners (MueLu or FROSch).

ShyLU also implements a distributed-memory linear solver based on the Schur complement method~\cite{ShyLUCore2014}. This hybrid approach combines direct and iterative solvers: each subdomain problem is solved in parallel using a direct solver, while the Schur complement is handled iteratively. The preconditioner for the Schur complement solver is computed using a probing approach or using a threshold-based dropping strategy. It has been developed to address the requirements of circuit simulation applications. The solver is also hybrid in terms of parallel computing, utilizing MPI+threads. However, it is not designed for GPU architectures, and therefore, it has not been implemented using Kokkos.

\subsection{Physics Block Operators and Preconditioners: Teko}
\label{sec:teko}

Many multi-physics problems are conveniently described by a block-structured matrix
which allows easy access to the underlying matrices representing single-field physics or coupling between different physical fields.
The Teko library~\cite{Cyr2016a} provides interfaces for operators and preconditioners that are constructed from such large physics-based sub-blocks.
These sub-blocks are Thyra operators which are themselves often implemented using Tpetra matrices.

Teko provides generic block preconditioning strategies such as $2\times2$ block-LU factorizations as well as block versions of Jacobi- and Gauss--Seidel-type methods for any number of blocks,
as well as commonly used approximate inverse strategies for the Navier-Stokes equation
such as SIMPLEC, LSC, and PCD preconditioners~\cite{CyrShadidEtAl2012_StabilizationScalableBlockPreconditioning}.
The Block-LU class offers the possibility to implement user-specific inverse approximation strategies,
for example using a sparse approximate inverse to construct the Schur complement~\cite{Firmbach2024a}.
More complicated multilevel hierarchies of block solvers can be generated via \code{Teuchos::ParameterList} objects.
Block preconditioners for first-order formulations of Maxwell's equations and Darcy flow are implemented in the Panzer package.
Teko's solvers can be registered in the Stratimikos interface for usage entirely driven by \code{Teuchos::ParameterList} objects.

\subsection{Eigensolvers: Anasazi}
Anasazi~\cite{Baker2009a} is an extensible and interoperable framework for large-scale eigenvalue algorithms.
This framework provides a generic interface to a collection of algorithms that are built upon abstract interfaces
that facilitate code reuse and extensibility.  Similar to Belos, Anasazi decouples the algorithms from the
implementation of the underlying linear algebra objects using traits mechanisms.  Concrete linear algebra adapters
are provided for Tpetra and Thyra, while users can also implement their own interfaces to leverage any existing matrix and vector representations. Any libraries that understand Tpetra and Thyra matrices
and vectors, like Belos and Ifpack2, may also be used in conjunction with Anasazi.  The suite of eigensolvers provided
by Anasazi includes locally optimal block preconditioned conjugate gradient (LOBPCG), block Davidson, Riemannian Trust-Region
(RTR), and block Krylov-Schur methods.  Recently, a family of trace minimization (TraceMin) methods and a
generalized Davidson method have been added to the suite of eigensolvers in Anasazi.

\subsection{Unified Solver Interface: Stratimikos}
\label{sec:Stratimikos}
The Stratimikos package offers a cohesive interface to various linear solvers and preconditioners within Trilinos, including Amesos2, Belos, FROSch, Ifpack2, MueLu, and Teko.
It requires that the matrix, right-hand side, and the solution vectors all support the Thyra interface.
Thyra also provides wrappers for Tpetra linear algebra.
Users can specify solver and preconditioner parameters via a \code{Teuchos::ParameterList}, which can be easily populated from an XML file.
Moreover, the package allows for the easy addition of new solver and preconditioner factories via adapters, with examples available in the \texttt{packages/stratimikos/adapters} directory.
Stratimikos does not support eigenvalue problems.
Anasazi has, however, a Thyra interface and can therefore be used with different linear algebra backends.

\section{Nonlinear Solvers and Analysis Tools}
\label{sec:nonlin_solve}
% !TEX root = main.tex

The packages included in this section provide the top-level algorithms for computational simulations and design studies.
These include nonlinear solvers, bifurcation tracking, stability analysis, parameter continuation, optimization, and uncertainty quantification.
A common theme of this collection is the philosophy of ``analysis beyond simulation,'' which aims to automate many computational tasks that are often performed by application code users by trial-and-error or repeated simulation. Tasks that can be automated include performing parameter studies, sensitivity analysis, calibration, optimization, and the task of locating instabilities.
Additional utilities for nonlinear analysis include abstraction layers and interfaces for application callbacks, as well as automatic differentiation tools that can provide the derivatives critical to analysis algorithms.

\subsection{Nonlinear Solvers: NOX, LOCA} \label{sec:nox}
NOX (Nonlinear Object-oriented Solutions) provides robust and efficient algorithms for solving systems of nonlinear equations.
NOX implements a number of Newton-based globalization techniques including line search~\cite{Pawlowski2006}, trust region~\cite{Pawlowski2006,Pawlowski2008}, and homotopy algorithms~\cite{Coffey2003}.
Additionally, it provides lower- and higher-order models including Broyden, Anderson acceleration \cite{Walker2011}, and tensor methods \cite{Bader2005}.
The algorithms have been designed for large-scale parallel inexact linear solvers using Krylov methods and support Jacobian-free Newton-Krylov variants.
The library interacts with application codes through the Thyra model evaluator interface.
NOX provides linear algebra abstractions for applications to use custom implementations with the algorithms.
All algorithms are implemented under base-class abstractions for the solver, direction, and line search objects.
Users can provide derived implementations for each of these base classes to insert new algorithms into the library.
NOX provides various stopping criteria, including absolute, relative, and weighted root mean square norms, as well as stagnation and NaN detection.
Applications can build custom stopping criteria within a tree-based logical structure and provide additional criteria through the \code{StatusTest} abstraction.
NOX also provides capabilities for continuation and stability analysis through the subpackage LOCA.

LOCA (Library of Continuation Algorithms)~\cite{Salinger2005}
provides techniques for computing families of solutions to nonlinear equations as well as methods for investigating their stability when these nonlinear equations define equilibria of dynamical systems.  It builds on the NOX nonlinear solver package to track solutions to sets of nonlinear equations as a function of one or more parameters (continuation).
Given an interface to NOX that defines the nonlinear equations, users only need to provide a method for setting the parameter values being varied.
LOCA provides several continuation methods, including pseudo-arclength continuation which allows tracking solution curves around turning points/folds.  Furthermore, LOCA has hooks to call the Anasazi eigensolver package to estimate leading eigenvalues of the linearization at each point along the continuation curve for linear stability analysis, including various spectral transformations highlighting eigenvalues in different regimes (e.g., largest magnitude, largest real).  Finally, LOCA implements equations augmenting the original nonlinear equations to locate and track bifurcation points where linear stability changes (e.g., turning point, pitchfork, and Hopf bifurcations) as a function of additional parameters. Examples of using LOCA for stability and bifurcation analysis include stability of a differentially heated cavity \cite{Salinger2002}, design of chemical vapor deposition reactors \cite{Pawlowski2001}, flow instabilities in counterflowing jets \cite{pawlowski_salinger_shadid_mountziaris_2006}, and ignition in laminar diffusion flames \cite{Shadid20061846}.

\subsection{Numerical Optimization: ROL}
Rapid Optimization Library (ROL) \cite{rol,ROL2022ICCOPT} is a package for numerical optimization. ROL brings an extensive
collection of state-of-the-art optimization algorithms to virtually
any application.
Its programming interface supports modern computational
hardware, including heterogeneous multi-core systems.
ROL has been used with great success for optimal control, optimal design,
inverse problems, image processing, and mesh optimization. Application areas
include geophysics, climate science~\cite{Perego2022}, structural
dynamics~\cite{AQUINO2019323,BUNTING2021107295}, fluid dynamics~\cite{Antil2023},
electromagnetics, quantum computing, hypersonics, and
geospatial imaging~\cite{Kouri2014}.

The key design, performance, and algorithmic features of ROL can be
summarized as follows:
\begin{itemize}
\item
\emph{Vector abstractions and matrix-free interface for universal applicability:}
Similar to other Trilinos packages that implement abstract numerical
algorithms, ROL's design is centered around an abstract linear algebra
interface, through the \code{ROL::Vector} class, which enables the use of any
ROL algorithm with any data type, such as the C++ \code{std::vector}, MPI-parallel
data structures (e.g., Tpetra, PETSc, HYPRE vectors), GPU-supporting data
structures (e.g., Kokkos, ArrayFire \cite{Yalamanchili2015}), etc.
In contrast to other vector implementations, ROL's abstract linear algebra
interface also enables the definition of custom vector dot products with duality pairings,
which are critical for performance in large-scale optimization with differential equations.
Finally, all ROL algorithms are matrix-free and rely on user-defined
applications of linear and nonlinear operators.  ROL's vector class design is
inspired by the Hilbert Class Library \cite{hcl}, the Rice Vector Library \cite{rvl},
and the RTOp framework \cite{rtop}, while ROL's operator interfaces for
simulation-based optimization, known as the SimOpt middleware,
are inspired by~\cite{Heinkenschloss1999}.
\item
\emph{Modern algorithms for unconstrained and constrained optimization:}
ROL features a large collection of well-established algorithms for smooth
optimization as well as several novel algorithms developed by the ROL team.
ROL categorizes optimization problems into four types based on the
presence of certain constraints: Type U (unconstrained), Type B (bound
constraints), Type E (equality constraints), and Type G (general constraints).
Additionally, each problem type can include linear constraints, which ROL can
reduce or eliminate algorithmically. The novel algorithms include augmented
Lagrangian methods for infinite-dimensional problems with general constraints
\cite{ALESQP} and matrix-free trust-region methods for convex-constrained
optimization \cite{Kouri2022}.
\item
\emph{Easy-to-use methods for stochastic and risk-aware optimization:}
ROL contains specialized interfaces and algorithms for stochastic optimization.
Optimization problems with stochastic and uncertain problem data pose challenges
in problem formulation and efficient numerical solution methods.  ROL's interface
for stochastic optimization enables the use of numerous modern risk measures,
probabilistic functions, and robust problem formulations. To approximate
stochastic problems~\cite{shapiro2021lectures}, ROL includes sample-based
approaches such as sample average approximation, stochastic approximation, and
adaptive sparse-grid quadrature~\cite{kouri2013trust,kouri2014inexact}. ROL
also includes specialized algorithms such as progressive hedging~\cite{rockafellar1991scenarios}
and the primal-dual risk minimization algorithm~\cite{kouri2022primal}.
\item
\emph{Fast and robust algorithms for nonsmooth optimization:}
ROL includes specialized algorithms to minimize the sum of a smooth function
and a nonsmooth, convex function.  This class of problems is ubiquitous in
computational science.  For example, nonsmooth sparsifying regularizers such as
the $\ell^1$-norm are common in data science and machine learning~\cite{fuhg2024extreme}.
These algorithms can be used to solve convex constrained optimization problems.
ROL's methods for this class of problems are built on the proximity operator
and include proximal gradient and Newton methods
\cite{beck2017first,kanzow2021globalized,ochs2014iPiano}.
As discussed in the subsequent bullet, ROL can also incorporate inexact
values and gradients for the smooth objective function term using trust regions~\cite{baraldi2023proximal}.
ROL also include an implementation of the trust bundle method~\cite{schramm1992bundle}
for solving general nonconvex, nonsmooth optimization problems.
\item
\emph{Trust-region methods for inexact and adaptive computations:}
ROL controls inexact function evaluations and exploits adaptive computations
and multi-fidelity models through its trust-region methods. Several of these methods
have been pioneered by the ROL team.  Their implementations typically require some
knowledge of error in the computations (i.e., an \emph{error estimate}).
For a comprehensive overview of problem formulations, algorithms with error control,
and their ROL implementations, see \cite{Kouri2018}.
\item
\emph{PDE-OPT application development kit for PDE-constrained optimization:}
To build and optimize models based on PDEs,
ROL offers an application development kit for finite element models, called PDE-OPT.
PDE-OPT is based on a modular design with three layers: local finite element computations,
which encode the physical equations governing the optimization problem;
global computations, which utilize Tpetra data structures;
and an interface to ROL, enabling efficient optimization.
To solve PDE-constrained optimization problems, users only need to implement the
local finite element layer of PDE-OPT (i.e., the evaluations of weak forms on
a mesh element); PDE-OPT automates all remaining steps.
\end{itemize}

In 2025, the development of ROL has moved to its own GitHub repository\footnote{\url{https://github.com/sandialabs/rol}}.
The ROL code is snapshotted into the Trilinos repository for convenience of Trilinos users.

\subsection{Automatic Differentiation: Sacado} \label{sec:sacado}

Sacado \cite{SacadoURL,phipps2012efficient,phipps2008large} provides forward and reverse-mode operator-overloading-based automatic differentiation (AD) tools within Trilinos.
Sacado's forward AD tools have been integrated with Kokkos and have demonstrated good performance on GPU architectures~\cite{phipps2022automatic}.
Sacado, along with its Kokkos integration, provides high-performance derivative capabilities to numerous Office of Science and NNSA (National Nuclear Security Administration) extreme scale applications, including: Albany for solid mechanics and land ice modeling~\cite{Salinger2016,MPASAlbany2018};
Charon for semiconductor device modeling~\cite{CharonUsersManual2020} and multiphase chemically reacting flows~\cite{Musson2009}; Drekar for computational fluid dynamics (CFD)~\cite{Sondak2021,Shadid2016}, magnetohydrodynamics~\cite{Shadid2016mhd}, and
plasma physics~\cite{Crockatt2022,Miller2019}; Xyce for electronic circuit simulation~\cite{xyceTrilinos,xycePCE}; and SPARC for hypersonic fluid flows~\cite{SparcValidation}.

\subsection{Uncertainty Quantification: Stokhos}

Stokhos~\cite{phipps2015stokhos,Phipps2016,phipps2014exploring} provides implementations of two intrusive uncertainty quantification strategies:
the intrusive stochastic Galerkin uncertainty quantification method~\cite{ghanem1990polynomial,ghanem2003stochastic} and the embedded ensemble propagation method~\cite{phipps2017embedded}.

Stokhos' implementation of the intrusive stochastic Galerkin uncertainty quantification method allows stochastic projections to be computed, such as polynomial chaos and generalized polynomial chaos expansions.
The implementation includes interfaces for forming the linear and/or nonlinear systems that follow from the stochastic Galerkin projection, as well as linear solver methods that can exploit the block structure of these systems.
In particular, stochastic Galerkin methods applied to PDEs employ a tensor product discretization between the spatial and stochastic discretizations that generates a two-level sparsity pattern in the resulting stochastic Galerkin operator.  This operator is traditionally organized in a block fashion where the outer sparsity pattern is dictated by the stochastic discretization while each block follows the sparsity pattern of the PDE spatial discretization.  To improve performance on architectures such as GPUs, Stokhos commutes this layout to obtain an outer sparsity pattern dictated by the spatial discretization where each block follows the stochastic discretization and leverages Kokkos hierarchical parallelism to implement calculations corresponding to the stochastic discretization in a scalable and efficient manner~\cite{phipps2014exploring}.
The implementation has been used in~\cite{constantine2014efficient} to efficiently propagate uncertainty in multiphysics systems by reducing the full system with a nonlinear elimination method.

The embedded ensemble propagation consists of propagating a subset of samples gathered into a so-called ensemble through the forward simulation at once.
It builds on~\cite{pawlowski2012automating} for automating embedded analysis capabilities. Stokhos defines an ensemble type: a data type that is able to store
the values of input, output, and state variables for every sample of an ensemble and can take advantage of SIMD (single instruction, multiple data) operations. This type can then be used within the Tpetra solver stack as an argument for its scalar type template parameter, allowing instantiation of Tpetra data structures on the ensemble type to propagate ensembles through supported linear solvers and preconditioners.
This approach results in reduced computation time in four ways: the sample-independent data and computation can be reused for every sample of an ensemble, memory access patterns can be improved,
operations on the ensemble type can be vectorized efficiently, and message passing costs can be reduced by sending fewer but larger messages.
However, the approach requires solvers and BLAS functions to be aware of the extra dimension associated to the ensemble; for example, GMRES for ensemble types~\cite{liegeois2020gmres} needs to monitor
the convergence of the individual samples in order to decide when to terminate based on whether every sample within the ensemble has satisfied the chosen stopping criteria.

\subsection{Nonlinear Analysis Tools: Piro}
Piro~\cite{osti_1231283} provides a simple and unified interface to nonlinear analysis tools in Trilinos, exposing capabilities for computing sensitivities, performing parameter continuation and bifurcation analysis, and solving constrained optimization problems. Piro wraps many of the common functionalities of Trilinos packages such as Belos, NOX, LOCA, Tempus and ROL, into a simple interface.
In particular, Piro implements main driver classes for:
\begin{itemize}
	\item \emph{Linear/nonlinear solvers and sensitivity analysis} for transient and nontransient problems. As an example, this capability can be used to compute the discrete solution of a partial differential equation depending on a parameter and the sensitivity of a quantity of interest with respect to this parameter. Sensitivities can be computed in a forward or adjoint fashion, the latter being preferable in the presence of high-dimensional parameters. To this end, Piro performs forward and adjoint solves, relying on Belos, NOX and Tempus for the solution of linear, nonlinear and transient problems.
	\item \emph{Constrained optimization problems} with linear/nonlinear equality constraints: Piro provides tools for transient and nontransient (snapshot) optimizations, featuring gradient-based reduced-space and full-space methods. This capability is used by applications such as Albany to solve large-scale PDE-constrained optimization problems \cite{Perego2022}. Piro interfaces with ROL for algorithms to solve constrained optimization problems.
\item \emph{Parameter continuation and bifurcation analysis.} The core capabilities for continuation and bifurcation analysis are provided through the package LOCA. The Piro driver integrates these functionalities with linear, nonlinear, and time integration solvers, enhancing flexibility and usability.
\end{itemize}
These driver classes share a similar interface based on the \code{Thyra::ModelEvaluator} and the \code{Teuchos::ParameterList} classes. Applications define the problems to be targeted (e.g., the equations to be solved, the parameters, the quantities of interests) by providing a concrete implementations of the \code{Thyra::ModelEvaluator}; see \cite{pawlowski2012automating,pawlowski2012automatingpart2}.

\section{Discretization Tools}
\label{sec:discretization}
% !TEX root = main.tex

This section describes Trilinos packages that provide tools for spatial and temporal discretization of integro-differential equations. Most discretization efforts in Trilinos have been devoted to implementing tools for mesh-based discretizations of PDEs with a focus on high-order finite elements. A notable exception is the research package Compadre (Section~\ref{sec:compadre}), which provides tools for meshless approximation of linear operators that can be used for the discretization of differential equations and for data transfer.
Trilinos' discretization packages have been adopted by many applications addressing a wide range of physics problems, including solid mechanics, earth system modeling, semiconductor devices modeling, and electro-magnetics. These applications have taken different approaches in adopting Trilinos mesh-based discretization tools. The less intrusive approach is the adoption of Intrepid2 tools to perform local finite element assembly. In this case, the application has to manage the global assembly, possibly using the \code{DoFManager} provided by Panzer, and FECrs matrix and vector structures provided by Tpetra.
A more intrusive approach is to additionally use the Phalanx package for managing dependencies of field evaluations in conjunction with the Thyra Model Evaluator and Sacado automatic differentiation, and possibly Tempus for time integration. This approach is particularly useful when developing solvers for complex multiphysics problems, because it allows easy re-use of computational kernels and automates the computation of Jacobians and sensitivities.
The most intrusive approach is to build the application around the Panzer package, which provides all of the above, plus the handling of linear and nonlinear solvers and integrated constrained optimization capabilities.
In the following, we neither include a description of the snapshotted packages Sierra ToolKit (STK) and Krino, which provide mesh and level-set tools, nor a description of the Shards package, which provides tools for mesh cell topology.

\subsection{Local Assembly: Intrepid2}
Intrepid2 provides interoperable tools for compatible discretizations of PDEs; it is a performance\hyp{}portable re-implementation and extension of the legacy Intrepid package \cite{bochev2012}. Intrepid2 mainly focuses on local assembly of continuous and discontinuous finite elements. It also provides limited capabilities for finite volume discretization.  Intrepid2 works on batches of elements (cells), and provides tools to efficiently compute discretized linear functionals (e.g., right-hand-side vectors) and differential operators (e.g., stiffness matrices) at the element level. Intrepid2 implements compatible finite element spaces of various polynomial orders for $H({\rm grad})$, $H({\rm curl})$, $H({\rm div})$, and $L^2$ function spaces on triangles, quadrilaterals, tetrahedrons, hexahedrons, wedges, and pyramids. It provides both Lagrangian basis functions and hierarchical basis functions \cite{fuentes2015}, and it implements performance optimizations (e.g., sum factorizations) exploiting the underlying structure of the problem (e.g., tensor-product elements or other symmetries).  The degrees of freedom of $H({\rm div})$ and $H({\rm curl})$ finite elements as well as high-order $H({\rm grad})$ finite elements depend on the global orientation of edges and faces and Intrepid2 provides orientation tools for matching the degrees of freedom on shared edges and faces. It also provides interpolation-based projection tools for projecting functions in $H({\rm grad})$, $H({\rm curl})$, $H({\rm div})$, and $L^2$ to the respective discrete spaces. Intrepid2 implements these capabilities through the following classes:
\begin{itemize}
\item \code{CellTools}: This class provides geometric operations in the reference and physical frames. This includes the computation of tangents and normals to edges and faces in the physical frame, the computation of the Jacobian of the reference-to-physical frame map, and other geometric computations.
\item \code{CubatureFactory}: This class provides quadrature rules (called \emph{cubatures} in Intrepid2) of various degrees of accuracy for approximating integrals over elements and their boundaries.
\item \code{Basis:} This is the base class that provides a common interface for functionalities related to finite element bases. Intrepid2 provides derived classes that implement this common interface for a variety of compatible finite element spaces. Each derived class implements the \code{getValues()} method that computes the values taken by the basis functions or their derivatives (e.g., the gradient for $H({\rm grad})$ functions, the curl for $H({\rm curl})$ functions) at a set of input points. The implementation of \code{getValues()} can be very different depending on the basis. Specific optimizations are available for tensor-product elements.  Additionally, there is a \code{BasisFamily} class with a convenience method, \code{getBasis()}, which constructs a basis depending on a template argument specifying the type of basis (e.g., hierarchical or nodal), the cell topology, the function space on which it is defined, and its polynomial degree.
\item \code{OrientationTools}: This class provides methods to orient the basis functions based on the global orientation of edges and faces, determined by the global numbering of the cell vertices. This is achieved by building a linear operator (a permutation for tensor-product elements) that encodes the orientation of a particular cell and applying that operator to the reference basis functions.
\item \code{ProjectionTools}: This class provides methods for interpolation-based projections of a given function into a compatible finite element space or between compatible finite element spaces~\cite{demkowicz2007}. The provided projections commute with the corresponding differential operators if the quadrature rules can exactly integrate the functions being projected. As an example, projecting an $H({\rm grad})$ function into the $H({\rm grad})$ finite element space and then taking its gradient gives the same result as taking the gradient of the function first and then projecting the gradient into the $H({\rm curl})$ finite element space.
\item \code{FunctionSpaceTools}: This class provides transformations of fields from the reference to the physical frame and back, computation of measures on edges, faces, and cells, scalar/vector/tensor multiplications and contractions for computing integrals.
\item \code{IntegrationTools}: This class provides integration methods that can take advantage of tensor product structures in basis values, providing mechanisms for performance\hyp{}portable, \emph{sum-factorized} assembly across $H({\rm grad})$, $H({\rm curl})$, $H({\rm div})$, and $L^2$ function spaces. In the future, we plan to provide similar interfaces to support matrix-free discretizations.
\end{itemize}
Intrepid2 makes use of Kokkos containers to enable memory layouts that are adapted to the computational platform. Intrepid2 also uses Kokkos for its core computational kernels, enabling threaded execution across a variety of architectures. The data types used by Intrepid2 are templated; it is therefore possible to propagate Sacado types through Intrepid2 to perform automatic differentiation. Current development of Intrepid2 focuses on providing efficient matrix-free discretizations to enhance efficiency on GPU architectures.

\subsection{Local Field Evaluation: Phalanx}
Phalanx is a local field evaluation library designed for equation assembly in PDE applications. The goal of Phalanx is to decompose a complex problem into a number of simpler problems with managed dependencies to support rapid development and extensibility of PDE codes \cite{Notz2012,pawlowski2012automating,pawlowski2012automatingpart2}. The data structures use Kokkos for performance portability. Through the use of template metaprogramming, Phalanx supports arbitrary user defined data types and evaluation types. This feature allows for simple integration of automatic differentiation tools via operator-overloaded scalar types from the Sacado package (Section~\ref{sec:sacado}). From a simple definition of equations, quantities such as Jacobians, Jacobian-vector products, Hessians, and parameter sensitivities can be evaluated to machine precision. These quantities can be used by other Trilinos packages for operations including Newton-based nonlinear solves, gradient-based optimization, constraint enforcement, and bifurcation analysis \cite{pawlowski2012automating,pawlowski2012automatingpart2}.

Phalanx uses a graph-based design to manage data dependencies. The runtime-defined directed acyclic graph (DAG) allows for rapid prototyping in a production environment where simple interfaces for analysts, flexible models/data structures, and integration of non-trivial third-party libraries are paramount. Phalanx is used in a number of large-scale parallel codes including Albany \cite{Salinger2016}, Charon \cite{CharonUsersManual2020}, Drekar \cite{Crockatt2022,Miller2019,Shadid2016mhd}, and EMPIRE \cite{BettencourtBrownEtAl2021_EmpirePic}.

In recent years, Phalanx has been extended to provide utilities for performance portability under automatic differentiation. For example, Phalanx provides tools for building and managing a Kokkos view-of-views on device without the use of unified virtual memory (UVM) and provides utilities for running virtual functions on device. In the future, these utilities may be separated into a separate package.

\subsection{Time Integration: Tempus}
Tempus (Latin, meaning time as in ``tempus fugit'' $\rightarrow$ ``time flies'') is Trilinos' time-integration package for advanced transient
analysis. It includes various time integrators and embedded
sensitivity analysis for next-generation code architectures.  Tempus
provides ``out-of-the-box'' time-integration capabilities, which
allows users to quickly and easily incorporate time-integration
capabilities in their applications and switch between various time
integrators depending on the simulation needs.  Additionally, Tempus
provides ``build-your-own'' capabilities, which allows applications
to incorporate various Tempus components to augment or replace
an application's transient capabilities. Other capabilities include
embedded error analysis, sensitivity analysis, and transient optimization
with ROL.

Tempus offers a general infrastructure for the time evolution of
solutions through a variety of general integration schemes.  Tempus
provides time integrators for explicit and implicit methods and for
first- and second-order ODEs.  It can be used from small systems of
equations (e.g., single ODEs for the time evolution of plasticity
models and multiple ODEs for coupled chemical reactions) to
large-scale transient simulations requiring exascale computing
(e.g., flow fields around reentry vehicles and magneto-hydrodynamics).

Tempus has several components that can be used in concert or
individually, depending on the needs of the application.
\begin{itemize}
  \item Integrators are the time-loop structure for time integration
  and provide several features, e.g., controlling the advancement of
  the solution, selecting the next time step size and handling the
  solution output.

  \item Time steppers are individual methods that advance the
  solution from one step to the next.  A variety of time steppers
  are available:
  \begin{itemize}
    \item \emph{Classic one-step methods} (e.g., forward Euler and trapezoidal methods),
    \item \emph{Explicit Runge-Kutta (RK) methods} (e.g., RK explicit 4 Stage),
    \item \emph{Diagonally Implicit Runge-Kutta (DIRK) methods} (e.g.,
    general tableau DIRK and many specific DIRK/SDIRK methods),
    \item \emph{Implicit-Explicit (IMEX) Runge-Kutta methods} (e.g., IMEX
    RK SSP2, IMEX RK SSP3, and general tableau IMEX RK methods),
    \item \emph{Multi-Step methods} (i.e., BDF2),
    \item \emph{Second-order ODE methods} (e.g., Leapfrog, Newmark methods,
    and HHT-Alpha),
    \item \emph{Steppers with subSteppers} (e.g., operator-split and
    subcycling methods).
  \end{itemize}

  \item Solution history is used to maintain the solution during
  time step failure, for solution restart/output, for interpolation of
  the solution between time steps, and to provide the solution for
  transient adjoint sensitivities.

  \item Time step control and strategies provide methods to select
  the time step size based on user input and/or temporal error
  control (e.g., bounding min/max time-step size, relative/absolute
  maximum error, and time step adjustments for output and checkpointing).
\end{itemize}

Additionally, Tempus has several mechanisms which allow users to
insert application-specific algorithms into Tempus components (e.g.,
through observers and creating derived classes).

\subsection{Finite Element Analysis: Panzer}
The Panzer package provides global tools for finite element analysis. The package is targeted for large-scale parallel implicit PDEs using continuous and discontinuous compatible finite elements as well as control volume finite elements. Panzer enables the solution of nonlinear problems by interfacing with several Trilinos linear and nonlinear solvers. It uses the Intrepid2 package for arbitrary-order finite element bases. It computes derivatives and sensitivities with the Sacado automatic differentiation (AD) package. Panzer achieves performance portability through the Kokkos programming model via Tpetra data structures.

Panzer is designed for multiphysics systems.
The assembly engine allows for different equation sets in each element block of the mesh and allows for mixed bases for degrees of freedom (DOF) within each element block
(e.g., mixed $H(\rm{div})$ and $H(\rm{curl})$ system for electro-magnetics).
The assembly tools can create a single fully coupled Jacobian matrix with all off-block dependencies (as a single \code{Tpetra::CrsMatrix}),
or it can create a blocked system, grouping sets of DOFs into separate explicit matrices.
This allows for physics-based block preconditioning strategies using the Teko package in Trilinos~\cite{Bonilla2023,Cyr2016a}
or MueLu's fully coupled AMG hierarchies for multiphysics systems~\cite{Ohm2022a}.
The assembly process relies on the Phalanx package for efficient assembly of the multiphysics systems.
Panzer additionally can wrap the assembly in a \code{Thyra::ModelEvaluator}, providing direct interfaces to the linear and nonlinear analysis packages in Trilinos.

Panzer can provide utilities for application codes to build on, or can be used as a high-level application framework. Important capabilities include the following.
\begin{itemize}
\item \code{DOFManager}: Panzer provides a stand-alone DOF manager class in the dof-mgr subpackage. Given a list of DOFs and their corresponding basis and element blocks, the DOF manager can provide the mapping from DOFs on the mesh entities to the entries in linear algebra objects such as residual vectors and Jacobian matrices. The DOF manager can return the objects required to build distributed Tpetra
maps and graphs for both uniquely owned global indices and ghosted indices used during assembly.
It provides the local indexing used during assembly as well. The DOF manager contains a mesh abstraction called a connection manager that provides information about mesh connectivity, global numbering of mesh topological entities, and element block groups. It is designed to support any underlying mesh database, allowing applications to use the DOF manager with any finite element application.
\item \code{STK\_Interface}: Panzer contains a concrete implementation of the connection manager API for the STK mesh database package in Trilinos. The \code{STK\_Interface} object wraps an STK mesh database. It provides a simple interface for accessing global indices of the mesh and can be used to read/write associated solution data to the database. It additionally supports SEACAS for writing mesh data to disk. The \code{STK\_Interface} also includes support for periodic boundary conditions.  This capability can match topological entities on periodic parallel distributed faces. Once matched, the DOFs are unified to enforce periodicity for the DOF manager. This capability is in the adapters-stk subpackage.
\item \emph{Linear Object Factory:} Panzer provides a linear object factory and linear object container designed to support parallel distributed assembly.
The returned containers hold either the linear algebra objects for a uniquely owned DOF map used for solving the linear system or a ghosted version of the linear objects that can be used for assembly. The containers support export and import operations between the unique and ghosted containers. These are used to simplify the assembly process and abstract the underlying Tpetra linear algebra objects.
The linear object factory can also create DOF gather and scatter functors for the corresponding (and possibly blocked) matrices. The gather and scatter operations are specialized on the assembly types, including residuals, Jacobians, and tangents.
Support for Hessians will be implemented as needed.
The capability is found in the disc-fe subpackage.

\item \emph{Worksets:} The Panzer assembly process provides workset containers to hold static data that can be reused in finite element computations. For example, when using a static mesh, these containers could hold the Jacobian, Jacobian inverse, and all finite element basis values at the quadrature points. The workset concept breaks the local elements on an MPI process into smaller blocks of uniform computations that can be dispatched to a CPU or GPU. The workset helps to control total memory use for temporaries in the Phalanx DAG. This capability is part of the disc-fe subpackage.
\item \emph{Assembly Kernels:} Panzer provides a number of assembly kernels that are optimized for use with Sacado AD data types. When assembling on GPUs, the assembly kernels were found to be an order of magnitude faster if the AD evaluation was parallelized over the internal derivative dimension \cite{phipps2022automatic}. This required using kernels with a \code{Kokkos::TeamPolicy} for finite element assembly, where the lowest level of parallelism is used internally by the Sacado AD type. When building Panzer for GPUs, the \code{DFad} hierarchic parallelism flag in Sacado should be enabled for kernel performance. The kernels can be found in the disc-fe subpackage.
\item Panzer provides an example miniapp for implicit electro-magnetics that is used for benchmarking linear solver performance and for acceptance testing of new high-performance computing systems. It demonstrates an $H(\rm{curl})$-$H(\rm{div})$ formulation for the electric and magnetic fields. It can be found in the MiniEM subpackage.
\end{itemize}

The Panzer package is intended to provide both low- and high-level tools for implicit finite elements discretizations. The high-level tools aggregate many Trilinos discretization and solver packages. While the high-level tools can be used as a rapid prototyping environment, the front-end is fairly complex to set up, as opposed to true rapid prototyping frameworks such as deal.ii~\cite{dealII95}, FEniCSx~\cite{BarattaEtal2023}, or Firedrake~\cite{FiredrakeUserManual}. Panzer is not user friendly in this regard, however, the examples and miniapps are a good starting point and can be quickly adapted to other physics applications. A number of performance portable applications are using Panzer tools at different levels of adoption; these include Albany \cite{Salinger2016}, Charon \cite{CharonUsersManual2020}, Drekar \cite{Crockatt2022,Miller2019,Shadid2016mhd}, and EMPIRE \cite{BettencourtBrownEtAl2021_EmpirePic}.

\subsection{Approximation of Linear Operators: Compadre}\
\label{sec:compadre}

The Compadre package provides tools for the approximation of linear operators (including point evaluation and derivatives), given the location of samples of a function over an unstructured cloud of points. The resulting stencils, when applied directly to samples of the function at these locations, provide an approximation of the linear operator acting on the function at the point(s) queried. This is useful for meshed and meshless data transfer applications (remap). Samples of the function at the specified locations can also be viewed as unknowns, in which case the solution returned by Compadre can be used as a stencil for meshless discretization of PDEs \cite{REBAR2024}.

The package uses generalized moving least squares (GMLS) for approximating functionals. GMLS generalizes classic moving least squares (MLS) in the sense that selecting point evaluations for the sampling functionals and the target operator in GMLS provides a traditional moving least squares reconstruction. However, GMLS offers significantly greater flexibility than MLS by allowing the evaluation of the target function at points to be replaced with \emph{sampling} functionals of a function (e.g., integrals of the function over local domains), and instead of approximating the target function, to approximate a target \emph{operator} (e.g., gradient, curl, divergence) of a function. A detailed description of GMLS can be found in \cite{mirzaei2012generalized,wendland2004scattered}.

This increased flexibility enables more advanced and unconventional remapping approaches. For instance, it is possible to use an average vector normal integral over edges (Raviart-Thomas type representation) or a cell-average integral (finite-volume representation) as the sampling functionals or the target operator \cite{gmd-15-6601-2022}. Compadre supports full space reconstruction in 1D, 2D, and 3D, and it also supports select sampling functionals and target operators on manifolds \cite{GROSS2020109340}.

Compadre's stencil generation involves independent problems to be solved in parallel at the team level with loops over the thread and vector level within each problem. This hierarchical parallelism is achieved with performance portability by using the Kokkos programming model and leveraging the batched QR with pivoting algorithm implemented in Kokkos Kernels.

Future plans include the implementation of other meshless methods like locally-supported radial basis functions.

\section{Trilinos Framework}
\label{sec:framework}
% !TEX root = main.tex

The Trilinos Framework product area provides supporting infrastructure for Trilinos users and developers. This infrastructure is largely not what Trilinos is well-known for but is essential for supporting all Trilinos capabilities.

\subsection{Build and Test Infrastructure}

The primary focus of the Trilinos Framework product relates to building, testing, and releases of Trilinos. This effort includes setting up and maintaining the general infrastructure for structured building of subsets of packages, automated testing, and maintaining and adding to the various testing configurations.

The Trilinos Framework infrastructure is built on top of the Tribal Build, Integration, and Test System (TriBITS~\cite{Bartlett2014}), which is built on top of the open-source tools CMake and CTest\footnote{\url{https://cmake.org}}.
The TriBITS framework allows building arbitrary subgraphs of dependent (Trilinos) CMake packages in one or more individual aggregated CMake projects (in any arrangement desired).
Each Trilinos/TriBITS package lists its direct (required and optional) dependent upstream packages, thus forming a package dependency graph.
The TriBITS framework uses this package dependency graph to automatically determine what indirect dependent internal packages must be enabled and processed (and built) and what external packages must be found.
TriBITS then orchestrates the processing of all of the required CMake code to find the needed external packages and configure and build (and optionally test and install) the selected set of internal packages.
This allows a large number of (Trilinos) CMake packages to be configured, built, and tested in a flexible and efficient manner.
In addition, TriBITS provides support for a number of advanced features that are not available in raw CMake/CTest including: eliminating a large amount of boiler-plate CMake code and avoiding common mistakes; enabling and testing all downstream packages given a set of enabled (i.e. modified) upstream packages; managing the enabling and disabling of tests based on various criteria; producing build and test results submitted to a CDash site on a package-by-package basis; producing reduced source tarballs for only a desired subset of enabled packages.
As of version 14.4 of Trilinos, TriBITS and Trilinos have been updated to allow packages using raw CMake to be integrated with just a few well-defined integration requirements. The TriBITS framework has allowed Trilinos to scalably grow in the number of packages and the complexity without undue burden on individual Trilinos developers and users.
An always up-to-date overview of the configurations used for automated testing of pull requests against Trilinos can be found in the source code repository\footnote{\url{https://github.com/trilinos/Trilinos/blob/master/.github/workflows/AT2.yml}}.

\subsection{Documentation Infrastructure}

The Trilinos Framework product is also responsible for maintaining the general infrastructure for the Trilinos website \footnote{\url{https://trilinos.github.io}}, wiki \footnote{\url{https://github.com/trilinos/Trilinos/wiki}}, and Doxygen documentation \footnote{\url{https://trilinos.github.io/documentation.html}}. It owns some of the documented workflows on these sites, such as the process for reproducing testing failures, but the Trilinos package developers own all of the documentation associated with their associated packages.

\subsection{Python Wrappers: PyTrilinos2}

PyTrilinos2 is a set of automatically generated Python wrappers for selected Trilinos packages including Tpetra, Teuchos and Thyra, and for exposing solver capabilities from Amesos2, Belos, Ifpack2, and MueLu through Stratimikos. In the future, the list of wrapped packages will be enlarged to provide users with more features and to enable efficient prototyping of new algorithms for developers.

\section{Trilinos Community}
\label{sec:community}
% !TEX root = main.tex

\subsection{Contributing to Trilinos}

Contributions to Trilinos can be offered through the standard GitHub pull request model. Proposed code changes are required to pass a set of tests as well as review and approval prior to be merged. Detailed instructions can be found in the contributing guidelines\footnote{\url{https://github.com/trilinos/Trilinos/blob/master/CONTRIBUTING.md}} in the source code repository.
Questions regarding the inclusion of new packages should be discussed with the Trilinos leadership team\footnote{\url{https://trilinos.github.io/team.html}}.

\subsection{Platforms for exchange among users or developers}

The community of Trilinos users and developers operates several forums for exchange and discussion. Technical discussions about the source code and its development happen within the Trilinos GitHub repository\footnote{\url{https://github.com/trilinos/Trilinos}}.
The \texttt{\#trilinos} channel within the Kokkos slack workspace\footnote{See \url{https://kokkos.org} for details.} provides a quick and accessible forum to ask questions.

For in-person exchange, the \emph{Trilinos User-Developer Group (TUG) Meeting} takes place at Sandia National Laboratories in Albuquerque every year. At TUG, all Trilinos users and developers can gather to inform themselves about recent progress and advances, and discuss current challenges and upcoming topics relevant to the entire Trilinos community.

The \emph{European Trilinos User Group (EuroTUG) Meeting} series\footnote{\url{https://eurotug.github.io}}
offers a platform for Europe-based users and developers of the Trilinos project.
EuroTUG offers Europe-based researchers and application engineers interested in the Trilinos project easy access to the Trilinos community, minimizing travel burdens by removing the need to travel to Albuquerque.
It includes tutorials, user presentations showcasing Trilinos applications, and developer updates on news and ongoing work.

\subsection{Embedding into other software initiatives}

Trilinos is a founding member of the High Performance Software Foundation\footnote{\url{https://hpsf.io}} (HPSF).
Established in 2024 under the Linux Foundation,
the HPSF aims to build, promote, and advance a portable software stack for HPC by fostering collaboration among industry, academia, and government entities.
As an initial technical project within the HPSF, Trilinos contributes its expertise in data structures for parallel computing, linear, nonlinear, and transient solvers,
as well as optimization and uncertainty quantification in support of HPFS' mission to enhance the HPC software ecosystem as a whole.

\section{Concluding remarks}
\label{sec:conclusion}
% !TEX root = main.tex

The evolution of the Trilinos project demonstrates its critical role in scientific and high-performance computing.
This update underscores Trilinos' dedication to maintaining relevance amidst the rapid development of HPC software frameworks and hardware architectures, and the demands of increasingly complicated multiscale multiphysics simulation code bases for tackling scientific and engineering problems.
By adopting the Kokkos ecosystem, expanding package functionalities, and embracing a modular structure,
Trilinos ensures both performance portability and adaptability to existing and emerging application domains.

The integration of data structures for distributed-memory paradigms, advanced linear and nonlinear solvers, discretization technology, and optimization tools highlights its versatility and impact across scientific and engineering disciplines.
Its community-driven development model fosters innovation while ensuring robust support for users and contributors alike.
Moving forward, the library's commitment to collaboration, scalability, and innovation positions it as a cornerstone of next-generation computational frameworks.

As the computational landscape evolves, Trilinos remains poised to address emerging challenges, leveraging its rich feature set and strong community foundation to drive progress in scientific discovery and technological innovation.

\section*{Acknowledgments}

Sandia National Laboratories is a multimission laboratory managed and operated by National Technology \& Engineering Solutions of Sandia, LLC, a wholly owned subsidiary of Honeywell International Inc., for the U.S. Department of Energy’s National Nuclear Security Administration under contract DE-NA0003525.

This paper describes objective technical results and analysis. Any subjective views or opinions that might be expressed in the paper do not necessarily represent the views of the U.S. Department of Energy or the United States Government.

This work was supported in part by the U.S. Department of Energy, Office of Science, Office of Advanced Scientific Computing Research, Scientific Discovery through Advanced Computing (SciDAC) Program through the FASTMath Institute.

This work was supported by the U.S.~Department of Energy, Office of Science, Office of Advanced Scientific Computing Research, Applied Mathematics program.

This work was supported by the Laboratory Directed Research and Development program at Sandia National Laboratories.

This work was supported in part by dtec.bw -- Digitalization and Technology Research Center of the Bundeswehr [project hpc.bw]. dtec.bw is funded by the European Union -- NextGenerationEU.

%%
%% The next two lines define the bibliography style to be used, and
%% the bibliography file.
\bibliographystyle{ACM-Reference-Format}
\bibliography{bibliography}

\end{document}